\documentstyle[aps,twocolumn,prb,epsfig]{revtex}

\begin{document}
\draft
\author{E. Nogueira Jr. and R. F. S. Andrade}
\address{Instituto de F\'{\i}sica\\
Universidade Federal da Bahia\\
Campus da Federa\c{c}\~{a}o\\
40210-340, Salvador, Bahia, Brazil}
\author{S\'ergio Coutinho}
\address{Laborat\'orio de F\'\i sica Te\'orica e Computacional,\\
Universidade Federal de Pernambuco,\\
50670-901, Recife, Pernambuco, Brazil}
\title{Multifractal properties of aperiodic Ising model:\\
role of the geometric fluctuations}

\wideabs{
\maketitle

\begin{abstract}
The role of the geometric fluctuations on the multifractal properties of the
local magnetization of aperiodic ferromagnetic Ising models on hierachical
lattices is investigated. The geometric fluctuations are introduced by
generalized Fibonacci sequences. The local magnetization is evaluated via an
exact recurrent procedure encompassing a real space renormalization group
decimation. The symmetries of the local magnetization patterns induced by
the aperiodic couplings is found to be strongly (weakly) different, with
respect to the ones of the corresponding homogeneous systems, when the
geometric fluctuations are relevant (irrelevant) to change the critical
properties of the system. At the criticality, the measure defined by the
local magnetization is found to exhibit a non-trivial $F(\alpha )$ spectra
being shifted to higher values of $\alpha $ when relevant geometric
fluctuations are considered. The critical exponents are found to be related
with some special points of the $F(\alpha )$ function and agree with
previous results obtained by the quite distinct transfer matrix approach.
\end{abstract}}

\section{Introduction}

The introduction of deterministic aperiodicity in the couplings and fields
has been intensively used to mimic of the effect of disorder on physical
homogeneous systems. This approach, opposed to the more conventional random
disorder distribution of coupling constants and fields, has been first used
in the analysis of electronic systems, and is now being used to investigate
the critical behavior of magnetic systems on Euclidean lattices, as well as
on fractal hierarchical lattices.

Hierarchical lattices have been largely used as a framework for analyzing
both homogeneous \cite{tsallis98} and quenched disordered magnetic systems 
\cite{southern77}. It is well known that they offer only crude
approximations for the critical properties of the corresponding homogeneous
systems on Euclidean spaces with the same fractal dimension, but they
provide an useful tool to investigate, via exactly solvable models, the
effect of both random and deterministic disorder. For instance, the short
range Ising spin glass model defined on the diamond hierarchical lattice has
been analyzed by means of a methodology that encompasses an exact real space
renormalization group decimation and an exact recursive procedure to
calculate the local magnetization of all lattice sites for a particular
realization of the disorder (sample)\cite{nogueira97}. A detailed numerical
investigation reveals that the multifractal properties of the local
Edwards-Anderson (EA) order parameter remains non-trivial far below the
critical temperature. This contrasts with previous investigations of the
homogeneous pure ferromagnetic Ising model on the same lattice, for which
the multifractal spectra of the normalized local magnetization survives only
at the critical temperature \cite{morgado90,morgado91}. Concerning the
critical behavior of the Ising spin glass model, it was possible to obtain
strong evidences of universal critical exponents and a unique value for the
temperature of the critical point when several probability distributions for
the quenched coupling constants are considered \cite{nogueira98,nogueira99}.
It is worth to mention that the values of the critical temperature and of
the critical exponents associated to the order parameter and correlation
length for Ising spin glass model on diamond hierarchical lattice with graph
fractal dimension $d_F=3$ are surprisingly close to those obtained by
numerical simulations for the model defined on cubic lattices \cite
{prakash97}.

In a recent publication, Luck \cite{luck93} proposed a heuristic criterion
to account whether the geometric fluctuations in coupling constants and
fields are relevant or irrelevant to alter (or not) the universality class
of the aperiodic systems with respect to the corresponding homogeneous one.
Luck's criterion plays the similar role of that due to Harris \cite{harris74}%
, that holds when quenched disorder is introduced in ferromagnetic systems.
In the latter, the critical behavior is changed provided the critical
exponent associated with the divergence of the specific heat of the
corresponding homogeneous system is positive. More recently, an exact
derivation of an analog of the Luck's criterion was presented for both the
Ising \cite{pinho98} and the Potts \cite{magalhaes98,haddad99} models,
defined on the family of diamond hierarchical lattice filled with aperiodic
interactions dictated by a generalized Fibonacci sequence. Using the
transfer matrix (TM) approach, Andrade \cite{andrade99} obtained exact
numerical results for the thermodynamic and critical properties of the
aperiodic ferromagnetic Ising model defined on the family of diamond
hierarchical lattices (hereafter DHL), corroborating the change of
universality class of the model when relevant geometric fluctuations are
considered accordingly to the Luck's like criterion properly adapted to cope
with hierarchical lattices. We notice that the exact scale invariance
symmetry is the crucial property of the DHL's underlying the methods which
link the thermodynamic functions of lattices with successive generations, as
occurs for the above mentioned real renormalization group scheme as well as
for transfer matrices approach \cite{andrade99}.

In this work we are mainly interested to investigate the role of geometrical
fluctuations on the multifractal properties of the local magnetization of
ferromagnetic Ising model on the DHL's. We focus our study on two distinct
models, both defined on DHL's with the same graph fractal dimension but with
distinct topologies leading to relevant and irrelevant fluctuations. To deal
with the multifractal properties of the local magnetization we apply the
methodology developed to study the multifractal spectra of the order
parameter of the homogeneous ferromagnetic Ising model \cite
{morgado90,morgado91,coutinho92} and the spin-glass Ising model \cite
{nogueira99} on the DHL's. As we will show, the caracter of the fluctuations
influences the nature of the changes observed in the singularity spectrum of
the local magnetization.

In section II, we describe the model Hamiltonian, present the sequences
governing the aperiodic interactions and discuss the relevance of the
geometric fluctuations for both lattices, accordingly with the appropriated
criterion presented in \cite{pinho98}. We also review the main features of
the flow diagram and the phase diagram of the aperiodic systems. Section III
is devoted to review the method to obtain the local magnetization and to
discuss the multifractal spectra for special realizations of the geometric
fluctuations. Finally we discuss and summarize our main conclusions in
section IV.

\section{The model Hamiltonian and the aperiodic interactions}

\begin{figure}
\leavevmode
\vbox{%
\epsfxsize=5cm
\epsfig{file=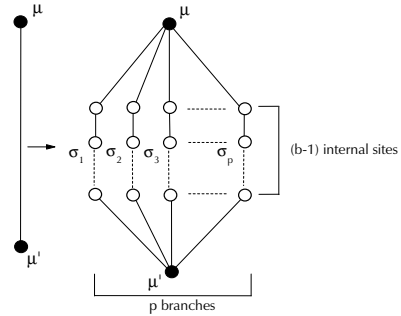,height=6cm,width=8cm}}
\caption{Basic unit of the general diamond hierarchical lattice $(b,p)$-DHL
with $p$ parallel branches each one with $b-1$ connected internal sites.}
\label{fig1}
\end{figure}

A general hierarchical lattice is constructed according to a recursive rule
where some bonds of basic unit are successively replaced by the basic unit
itself. The general basic unit of the diamond hierarchical lattices $(b,p)$%
-DHL, is formed by two roots sites connected by a set of $p$ parallel
branches, each one containing a series of $b$ connected bonds. A given
generation is obtained by replacing all bonds of the previous generation by
the basic unit, as sketched in Figure \ref{fig1}. For this family of
hierarchical lattices, the graph fractal dimension is given by $d_F=1+\log
p/\log b$, while its size, number of sites and the number of bonds are
respectively given by $L=b^N$, $N_S=(b-1)p[(bp)^N-1]/(bp-1)$ and $N_B=(bp)^N$%
, $N$ being the number of generations.

The reduced Hamiltonian for the system is

\begin{equation}  \label{eq1}
-H/k_BT=\sum\limits_{\left\langle i,j\right\rangle }K_{ij}\sigma _i\sigma
_j\ ,
\end{equation}
where $\sigma _i=\pm 1$ are the Ising variables and the sum runs over all
pairs of nearest neighbors spins $\left\langle i,j\right\rangle $ , $%
K_{ij}=J_{ij}/k_BT$ are the corresponding coupling constants, $k_B$ is the
Boltzmann constant and $T$ is the absolute temperature.

In this paper, we consider geometric fluctuations introduced by the
generalized Fibonacci inflation rule expressed by

\begin{equation}  \label{eq2}
(A,B)\rightarrow (AB^{b-1},A^{b})\ .
\end{equation}

When applied to define the coupling constants of the hierarchical lattice $%
(J_{ij}=J_A;J_B)$, the above rule corresponds to the following procedure: $%
(a)$ a bond with coupling constant $J_A,(J_A>0)$ should be replaced by a
basic unit with $p$ parallel branches each one containing the first bond
with coupling constant $J_A$ and the next $(b-1)$ ones with coupling
constants $J_B$; $(b)$ a bond $J_B,(J_B>0)$ should be replaced by a basic
unit with all bonds with coupling constants $J_A$. This correspondence is
displayed in Figures 2a and 2b, for the cases of the lattices with $p=b=2$
and $3$, respectively. After N steps, the length of the aperiodic sequences
on which we base the present definition of the geometric fluctuations is
given by $b^N$, which corresponds to the number of bonds within each one of
the shortest paths connecting the two root sites and measures the size of
the lattice. We also notice that there are $p^N$ of such shortest paths, all
of them with no common bonds and showing the same aperiodic sequence.
Therefore the systems can be regarded as formed by identical layers with the
bonds $J_A$ and $J_B$ arranged accordingly to the sequence (\ref{eq2}). The
number of distinct letters or coupling constants within two subsequent
generations of the sequence are related by the {\em substitution} matrix $M$
given by 
\begin{equation}
M=\left[ \matrix{1 & b \cr b-1 & 0\cr}\right] ,  \label{eq3}
\end{equation}
whose eigenvalues are $\lambda _1=b$ and $\lambda _2=(1-b)$. The wandering
exponent which measures, the fluctuation in the distribution of elements
within the sequence, is defined as 
\begin{equation}
\omega =\frac{\log \left| \lambda _2\right| }{\log \lambda _1}.\ 
\label{eq4}
\end{equation}
The sequences defined by (\ref{eq2}) are said to be non-Pisot if $\omega >0$
($b>2$), which means that the fluctuations are unbounded, whereas the case $%
\omega =0$ ($b=2$) corresponds to marginal case. On the other hand, for the
Ising model on DHL with aperiodic interactions arranged by sequences such
that $\lambda _1=b$, the geometric fluctuations are said to be {\em relevant 
} provided 
\begin{equation}
\omega >\omega _c=1-1/\nu ,  \label{eq5}
\end{equation}
$\nu $ being the critical exponent associated with the correlation length of
the homogeneous system \cite{pinho98}.

In the present work we consider two models: model 1 is defined on a $(2,2)$%
-DHL while model 2 is defined on a $(3,3)$-DHL. 
\begin{figure}
\leavevmode
\vbox{%
\epsfxsize=4cm
\epsfig{file=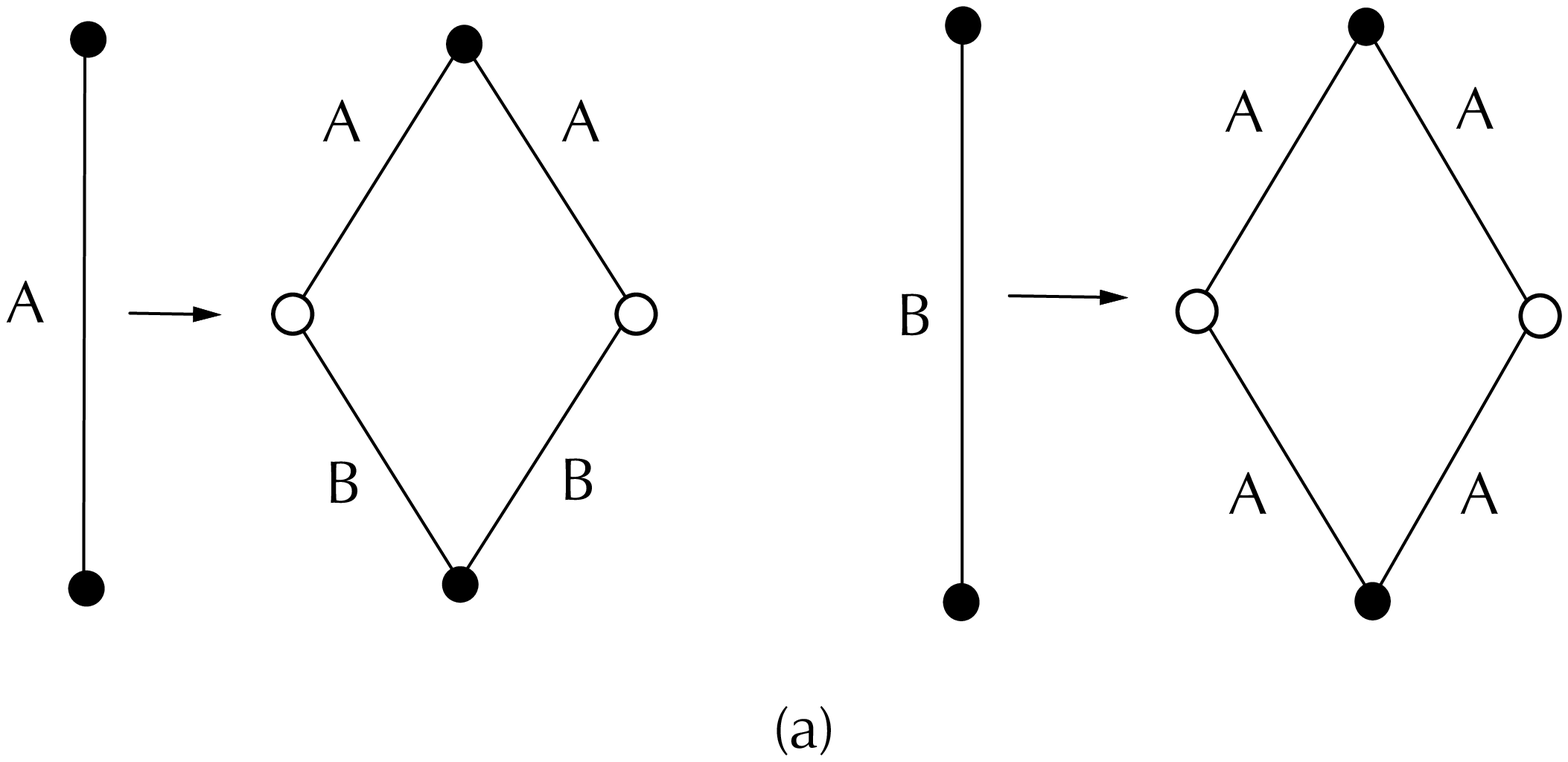,height=6cm,width=7cm}}
\leavevmode
\vbox{%
\epsfxsize=4cm
\epsfig{file=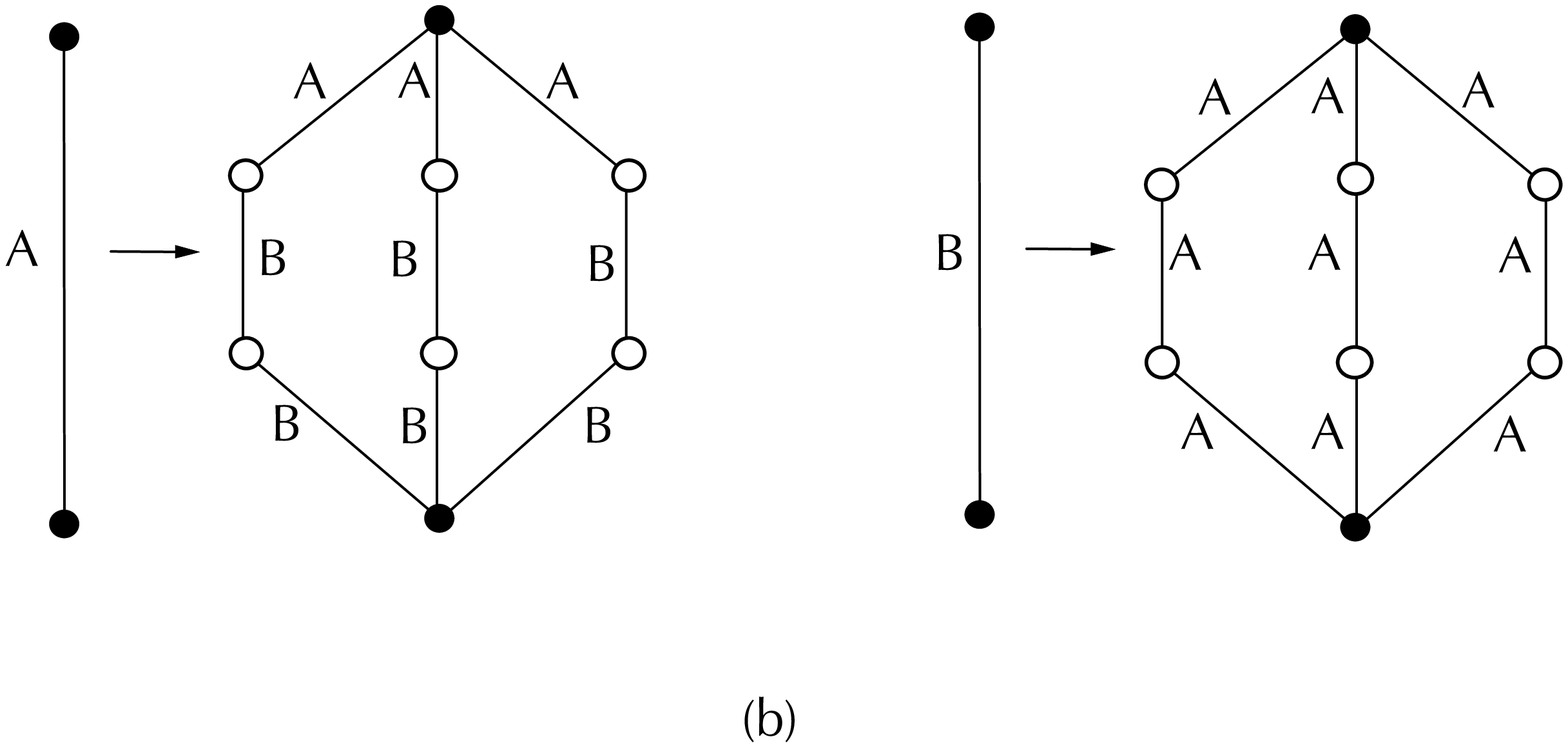,height=6cm,width=7cm}}
\caption{Basic units of the (a) $(2,2)$-DHL and (b) $(3,3)$-DHL indicating
the distribution of couplings $J_A$ and $J_B$ accordingly with the
generalized Fibonacci sequences.}
\label{fig2}
\end{figure}
The corresponding lattices
have been choosen to have the same fractal dimension ($d_F=2$), but the
geometric fluctuations are respectively irrelevant and relevant respectively
accordingly to eq.(\ref{eq5}), as has been demonstrated by Pinho et al \cite
{pinho98}. In the rest of this section we review the basic properties of the
renormalization flow of the model Hamiltonian in these lattices.

Consider a general $(b,p)$-DHL with the coupling constants defined by the
inflation rule given by (\ref{eq2}). Within the real space renormalization
group scheme, the decimation procedure is carried on by partial tracing on
the spins introduced in the last generation. This leads to exact scaling
relations for the coupling constants $J_A$ and $J_B$, given respectively the
following renormalization equations: 
\begin{eqnarray}
t_A^{\prime } &=&\tanh [p\tanh ^{-1}(t_At_B^{b-1})]  \label{eq6} \\
t_B^{\prime } &=&\tanh [p\tanh ^{-1}(t_A^b)]  \nonumber
\end{eqnarray}
where $t_x=\tanh K_x,$ $\ x=A$ or $B$.

The renormalization flow diagram in the $(t_A,t_B)$ space has been carefully
studied in \cite{pinho98,haddad99}. The equations (\ref{eq6}) have three
fixed point solutions, all of them located along the manifold $t_A=t_B$: $%
t_\infty ^{*}=0$, $t_0^{*}=1$, and $t_c^{*}$. Therefore, they are also the
solutions of the corresponding homogeneous system. The first two solutions
refer to stable fixed points corresponding to the infinite (paramagnetic)
and zero (ferromagnetic) temperature phases respectively while the third one
is associated with the critical point governing the transition of the
corresponding homogeneous system. For the particular models considered here,
these critical point solutions are exact and given by, 
\begin{eqnarray}
t_{c1}^{*} &=&\frac 13(a_c-2/a_c-1)=0.543689...\text{ for }p=b=2  \label{eq7}
\\
t_{c2}^{*} &=&\frac 12(\sqrt{5}-1)=0.618033...\text{ for }p=b=3  \nonumber
\end{eqnarray}
where $a_c=(3\sqrt{33}+17)^{\frac 13}$. As shown in \cite{pinho98}, the
renormalization flow space has two regions corresponding to the basins of
attraction of the stable fixed points. The frontier of these regions,
containing the critical point of the homogenous system, determines the phase
diagram. However, as pointed out in \cite{pinho98,magalhaes98,haddad99},
linearization about the non-trivial fixed point indicates that it is a
saddle-point for lattices with $b=2,(\omega =0<\omega _c)$, while it becomes
fully unstable for lattices with $b>2,(\omega >\omega _c)$. Therefore, for
model 1 the critical behavior should be governed by the fixed-point of the
homogeneous system and the geometric fluctuations are {\em irrelevant} with
respect to change the universality class of the model. On the other hand,
for model 2, the non-trivial fixed point is fully inaccessible indicating
that the geometric fluctuations are indeed relevant to change the critical
behavior. To grasp the critical behavior of the relevant aperiodic model one
should explore further the flow diagram given by eqs.(\ref{eq6}). This was
already done by Haddad et al \cite{haddad99} investigating the Potts model
on the $(3,2)$-DHL. Actually, whenever the fixed-point becomes fully
unstable, they found the appearance of a two-cycle attractor associated with
the novel critical behavior. For the present model 2, this two-cycle
attractor is given by coordination points{\bf \ }($%
t_{A1}^{*}=0.95116...,t_{B1}^{*}=0.46135...$) and ($%
t_{A2}^{*}=0.54822...,t_{B2}^{*}=0.99915...$) \cite{tesehaddad99}.

In Figure 3, we show the $(t_A,t_B)$ flow diagram for the $p=b=3$ case. 
Note that thetwo-cycle attractor is a saddle-point being unstable towards 
the trivial stable fixed-points and stable in the perpendicular direction. 

\begin{figure}
\leavevmode
\vbox{%
\epsfxsize=4cm
\epsfig{file=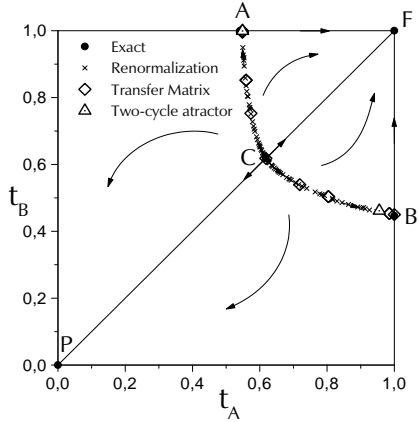,height=6cm,width=8cm}}
\caption{ The flow diagram in the $t_A\times t_B$ space for the real 
space renormalized couplings for model 2. The arrows indicates the 
flow of the second iterates. F and P denote the stable fixed-points 
corresponding to the ferromagnetic and paramagnetic phases. C labels
the fully unstable fixed-point. $\triangle$ locates the two-cycle 
attractor and $\diamond$ the critical points obtained by the transfer 
matrix method. $\times$ labels the critical points obtained by 
inverting the renormalization equations. A and B give the frontiers of 
the basin of attraction within the manifolds $t_B=1$ and $t_A=1$ 
respectively}
\label{fig3}
\end{figure}

The frontier between the basin of attraction of the stable fixed-points 
contains both the two-cycle attractor and the fully unstable fixed-point. 
This frontier intersects the lines $\ t_B=1$ and $t_A=1$ in two points, 
respectively $A$
and $B$, and it manifold is associated with the critical temperatures of the
model for different choices of the ratio $R=J_B/J_A$. 

The flow diagram can
be numerically obtained by inverting eqs.(\ref{eq6}) taking into account
that the inverse equations share the same trivial, non-trivial fixed points
and two-cycle attractor as eqs.(\ref{eq6}), but with inverted unstable to
stable manifolds and vice-versa. So, the iterates of these inverted
equations starting close to the critical fixed point are now pushed towards
line $ACB$, so that a relation $t_B=g(t_A)$ can be numerically obtained
ending to the phase diagram. 

In Figures 4a and 4b, we draw such relation in
the $(T/J_A,R)$ phase diagrams of both models, indicating the critical
temperature obtained for some special values of the ratio $R$ by using the
TM formalism \cite{andrade99}, as well as the corresponding exact values for
the homogeneous models ($R=1$). 
The critical exponent $\nu $ associated with
the correlation length of the corresponding homogeneous systems can be
exactly calculated as 
\begin{equation}
\nu =\frac{lnb}{lnr_c}  \label{eq8}
\end{equation}
\begin{figure}
\leavevmode
\vbox{%
\epsfxsize=4cm
\epsfig{file=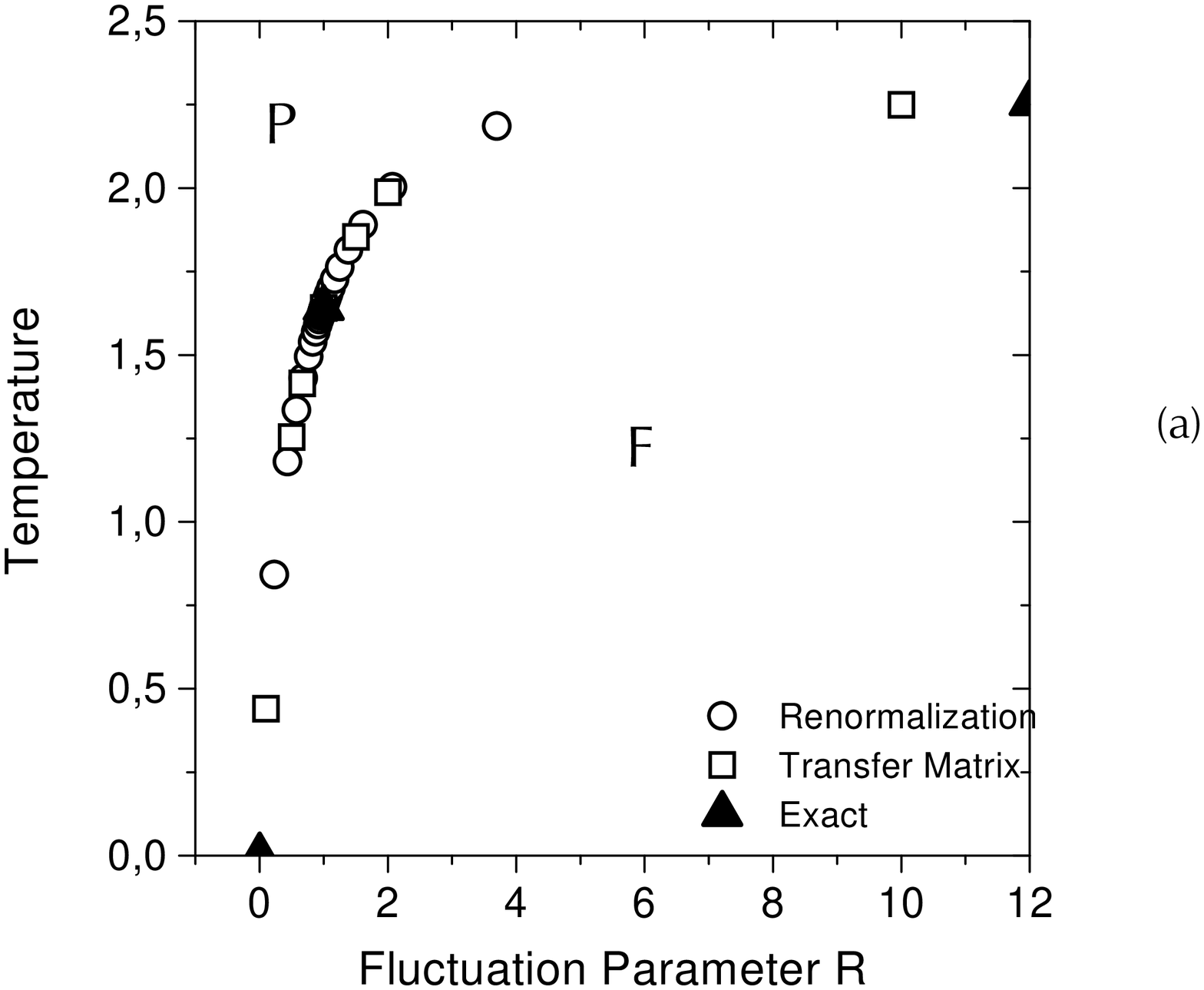,height=6cm,width=8cm}}
\leavevmode
\vbox{%
\epsfxsize=4cm
\epsfig{file=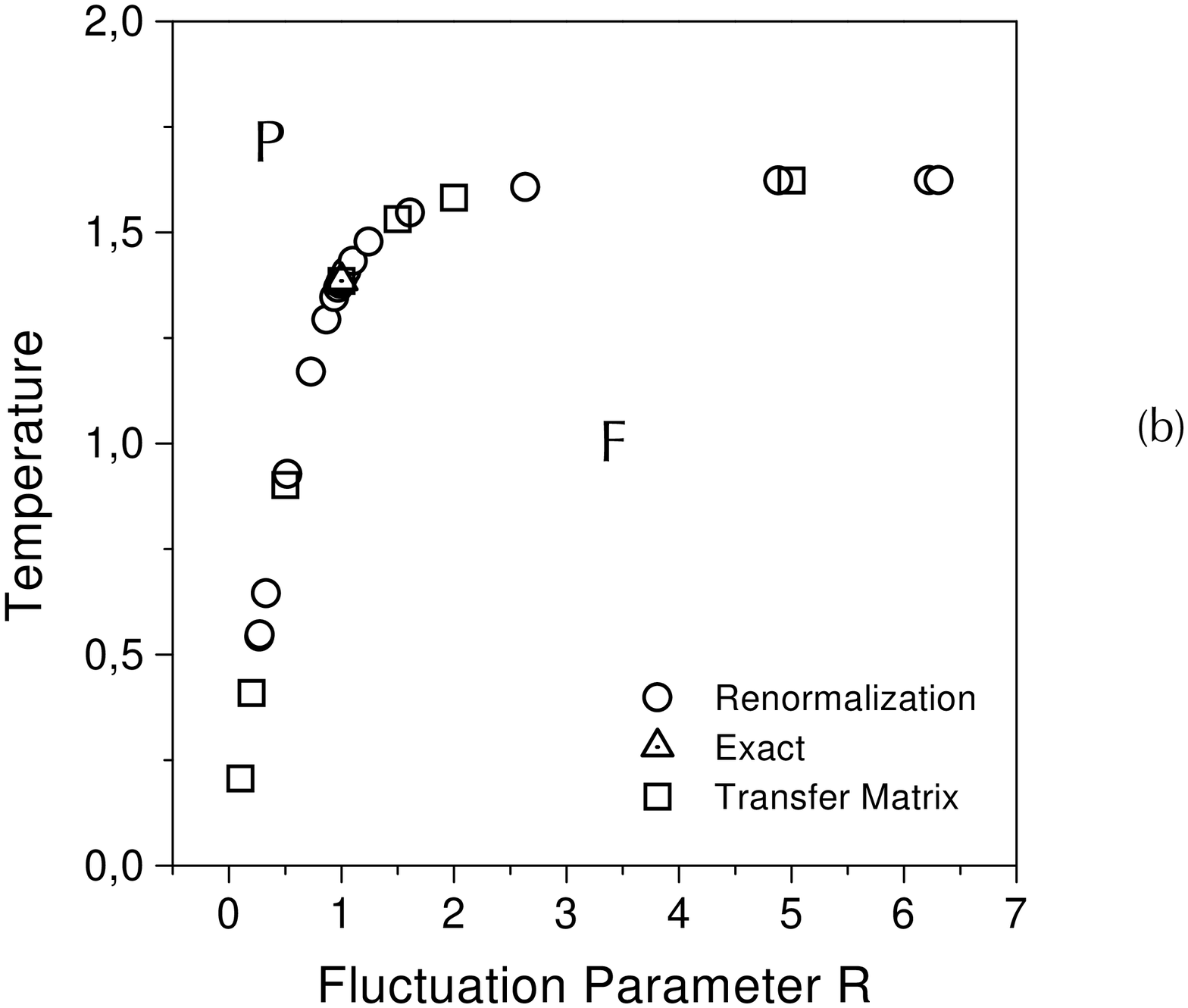,height=6cm,width=8cm}}
\caption{Phase diagram $K_BT/J_A\times J_B/J_A$ for (a) model 1, (b) model
2. $\bigcirc $ ($\bullet $) indicate numerical (exact) values obtained with
the present methodology , while $\triangle $ indicate values obtained by
transfer matrix method. $\triangle $ label the exact values for the
homogeneous systems}
\label{fig4}
\end{figure}
\noindent
where 
\begin{equation}
r_c=\left| {\frac{dt^{\prime }(t)}{dt}}\right| _{t=t_c^{*}}  \label{eq9} \\
={\frac{4bpt_c^{*b}(1+t_c^{*b})^{p-1}(1-t_c^{*b})^{p-1}}{%
((1+t_c^{*b})^p+(1-t_c^{*b})^p)^2}},
\end{equation}
and $t^{\prime }(t)$ is given by eqs.(\ref{eq6}) when $t_A=t_B=t$. 
\begin{table}
\caption{ Exact values for $\nu $, $\omega$ and $\omega _{c}$ for models 1
and 2.}
\label{tab1}
\begin{tabular}{ccc}
& model 1 & model 2 \\ \hline
$\nu$ & 1.338265788... & 1.779416044... \\ 
$\omega$ & 0 & 0.630929753... \\ 
$\omega_{c}$ & 0.252764279... & 0.438017880...
\end{tabular}
\end{table}

In Table \ref{tab1}, we show the values of $\nu $, $\omega $ and 
$\omega _c$ for models 1 and 2. It corroborates that for model 2 
we have $\omega >\omega _c$ indicating that the geometric fluctuation
should be {\em relevant} for this model.

\section{The Local Magnetization}

The method of evaluation the local magnetization is based on the assumption
that the model Hamiltonian for a lattice with $N$ generations is equivalent
to a reduced effective Hamiltonian of a single basic unit introduced in the $%
N^{th}$ step plus effective fields acting on the spins of its external sites
and an effective interaction coupling these spins. This assumption, which
has been proved to be formally correct for the homogeneous system, also
holds for the present model, since no special condition is imposed to the
coupling constants of the Hamiltonian. These unknown local effective
interaction and fields represent the influence of the remaining lattice spin
couplings transmitted by the external spins of that basic unit. The local
magnetization of both internal and external sites of the basic unit of the
effective system are calculated as a function of coupling constants $(J_A$
and $J_B)$ and the unknown effective coupling and fields. Eliminating the
unknown variables we end up to recursive relations for the local
magnetization of the internal sites in terms of the corresponding values of
the root sites. Now, considering a $N$ generation lattice and successively
decimating the spins up to the first generation (basic unit) one can
recursively calculate the local magnetization by choosing appropriated
values for the magnetization of root sites, accordingly with the
configuration of the phase of the corresponding fixed point. In \cite
{morgado90} the details of the calculations of the recursive equations for
the uniform ferromagnetic case is presented, while in \cite{nogueira97}
these equations were generalized for basic unit with arbitrary interactions
in order to investigate the short range Ising spin glass model. The
recursive equations for the latter can be straightforwardly applied to
deterministic aperiodic Ising models expressing the site magnetization of
internal sites $\left\langle \sigma \right\rangle $ in terms of the
corresponding values for the external sites. The recursive expression for
model 1 is given by : 
\begin{equation}
\left\langle \sigma \right\rangle =A_1\left\langle \mu \right\rangle
+B_1\left\langle \mu ^{\prime }\right\rangle  \label{eq10}
\end{equation}
where 
\begin{eqnarray}
A_1 &=&\frac{t_A(1-t_At_B)}{(1-t_A^2t_B^2)},  \label{eq11} \\
B_1 &=&\frac{t_B(1-t_At_B)}{(1-t_A^2t_B^2)},  \nonumber
\end{eqnarray}
for the $(AB)$ non-homogeneous basic unit while 
\begin{equation}
A_1=B_1=\frac{t_A}{(1+t_A^2)}  \label{eq12}
\end{equation}
for the $(A,A)$-homogeneous basic unit, as specified in Figure 2a.

On the other hand, for model 2 we have two sets of recursive equations
obtained for each one of the basic units shown in Figure 2b. Each set has
two equations corresponding to the internal sites of each basic unit. For
the non-homogeneous basic unit (ABB) these equations are given by 
\begin{equation}
\left\langle \sigma _i\right\rangle =A_i\left\langle \mu \right\rangle
+B_i\left\langle \mu ^{\prime }\right\rangle \hspace{3cm}(i=1,2),
\label{eq13}
\end{equation}
where 
\begin{eqnarray}
A_1 &=&\frac{t_A(1-t_B^4)}{(1-t_A^2t_B^4)},  \label{eq14} \\
B_1 &=&\frac{t_B^2(1-t_A^2)}{(1-t_A^2t_B^4)},  \nonumber \\
A_2 &=&\frac{t_At_B(1-t_B^4)}{(1-t_A^2t_B^4)},  \nonumber \\
B_2 &=&\frac{t_B^2(1-t_A^2t_B^2)}{(1-t_A^2t_B^4)},  \nonumber
\end{eqnarray}
while for the homogeneous basic unit (AAA) they are expressed by 
\begin{equation}
\left\langle \sigma _i\right\rangle =C_i\left\langle \mu \right\rangle
+D_i\left\langle \mu ^{\prime }\right\rangle \hspace{3cm}(i=1,2),
\label{eq15}
\end{equation}
where 
\begin{eqnarray}
C_1=D_2= &&\frac{t_A(1-t_A^4)}{(1-t_A^6)},  \label{eq16} \\
D_1=C_2= &&\frac{t_A^2(1-t_A^2)}{(1-t_A^6)}.  \nonumber
\end{eqnarray}

To evaluate the correct value of the local magnetization $m_i=\left\langle
\sigma _i\right\rangle $ of each site, it is necessary to keep track of its
local interaction configuration, i.e., to know with which choice of $J_A$
and $J_B$ a given inner spin interacts with its root sites. In order to
breakdown the global up-down symmetry and obtain non-zero values for the
spontaneous magnetization, it is enough to apply a field on a given site of
the lattice and let this local field equal zero at the end of calculation.
However, within the present methodology this breakdown is easily done by
assuming boundary conditions to root sites accordingly with the
corresponding configuration of the stable fixed-point phase. This latter
procedure is followed in the present work.

Now we focus our attention to model 2 looking over the effects of the
relevant geometric fluctuations with regard to the homogeneous pure model on
the same lattice. First, we consider the local magnetization of the sites
along one of the shortest path connecting the root sites. These values are
assigned to points within the interval $[0,1]$, since the DHL is a fractal
graph where the bonds mean ''chemical distances'' without any geometric
sense. The resulting plot gives a representative picture of the local
magnetization distribution, hereafter called magnetization profile, since
all shortest path are equivalent as far the DHL graph symmetry is concerned.
Figure 5a shows the magnetization profile for the homogeneous model 2 at the
critical temperature of the corresponding saddle-point fixed-point, and
Figures 5b and 5c show the profiles for the aperiodic model 2 at the
critical temperatures of each one point of the two-cycle attractor for
(which correspond to fluctuation parameter values $R_{c1}=J_B/J_A=0.2706...$
and $R_{c2}=J_B/J_A=6.3104...$. For the first profile, in any of the
renormalization steps, the coefficients (\ref{eq14},\ref{eq16}) have their
values fixed by the corresponding value at the critical point. For the other
two profiles these coefficients alternate between the corresponding values
at the two-cycle attractor. For values of $J_A$ and $J_B$ other than those
corresponding to the critical sets, the present methodology produces the
local magnetization as function of the temperature, but the above mentioned
coefficients evolve under the renormalization process. In any case, the
magnetization of the global root sites are held fixed with initial values
equal to one, which corresponds to the ferromagnetic configuration.

All profiles shown display many singularities and are, in some sense, self
similar. However those of the aperiodic system clearly do not show the same
symmetries as the one from the homogeneous model ($R=1$). As pointed out in 
\cite{morgado90,morgado91}, for the homogeneous model the pattern symmetries
are solely related with the topologic symmetry of the lattice, imposed by
the distribution of coordination numbers along the profile. However, for the
aperiodic systems, the self similar symmetry of the lattice topology is
superimposed by symmetries of the aperiodic sequence defining the
distribution of interactions, so that the resulting pattern expresses the
influence of both symmetries. We emphasize that this feature does not depend
on the relevant-irrelevant character of the fluctuations, and is observed
also for the aperiodic situation of model 1. It is worth to comment that for
the spin-glass model the symmetry induced by the lattice topology is
completely washed out by the random quenched disorder of the coupling
constants \cite{nogueira97}.

The profiles in Figures 5b and 5c are, at first sight, rather different, but
a close analysis uncover the fact that they share the same structure. We
note also the presence of two different scales corresponding to the regions
where the letters $A$ and $B$ are more abundant. As the values of $J_A$ and $%
J_B$ are rather distinct for the two points of the cycle, they enhance or
depress the values of the local magnetization in these regions, leading to
the two different shapes. 

It is important to observe that both profiles were
obtained after eight generations, so that, in the flow diagram, the
trajectories start and end at the same point of the cycle. 

If we consider an
odd number of generations the form of the profiles would be reversed,
indicating that the profile does not converge to a single form, but rather
oscillates between two distinct patterns which reflect the properties of
each of the points in the cycle. The same behavior is observed for every
profile, evaluated with any choice of values for $J_A$ and $J_B$, at the
corresponding value of $T_c$.

\begin{figure}
\leavevmode
\vbox{%
\epsfxsize=6cm
\epsfig{file=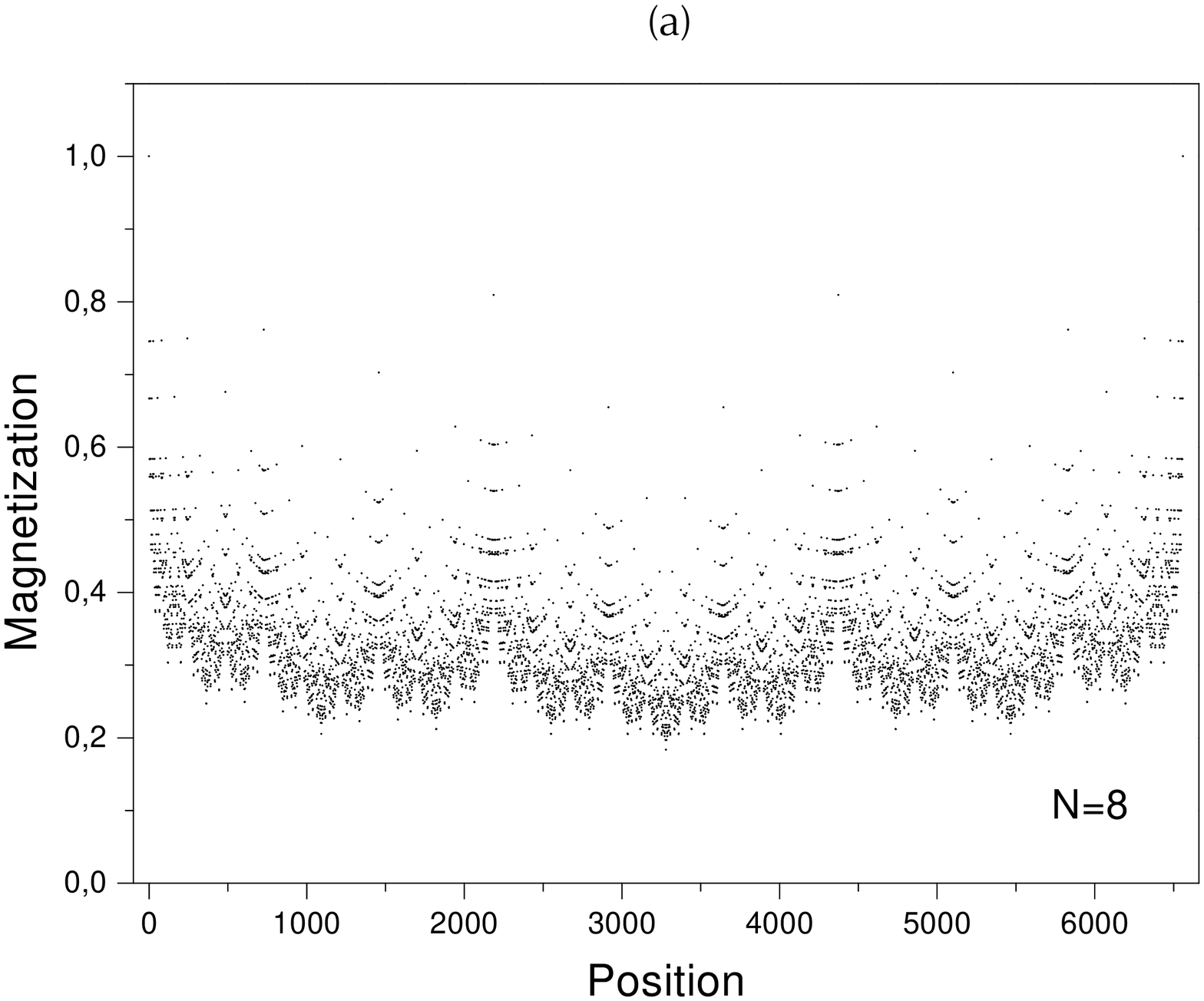,height=6cm,width=7cm}
\leavevmode
\vbox{%
\epsfxsize=6cm
\epsfig{file=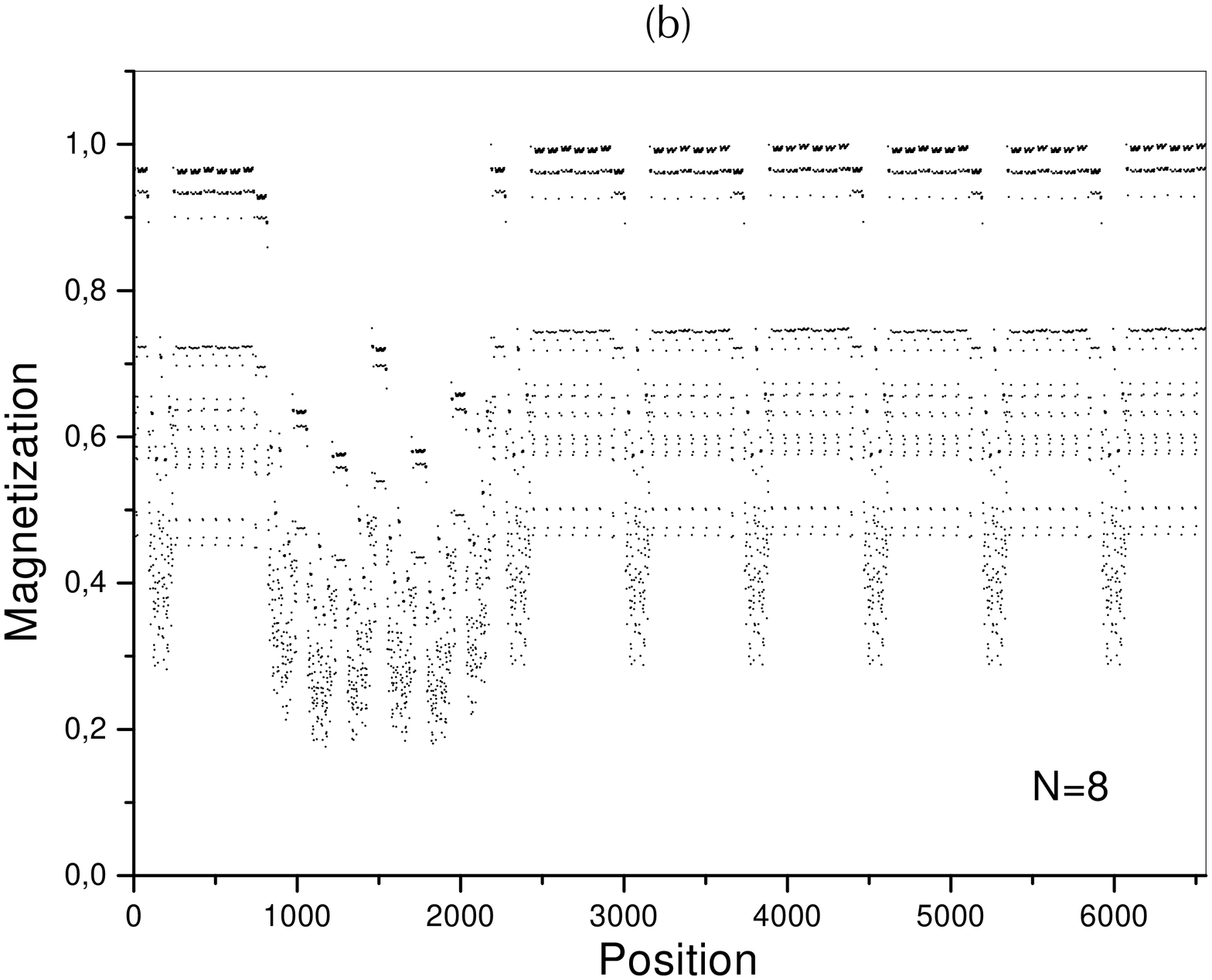,height=6cm,width=7cm}}
\leavevmode
\vbox{%
\epsfxsize=4cm
\epsfig{file=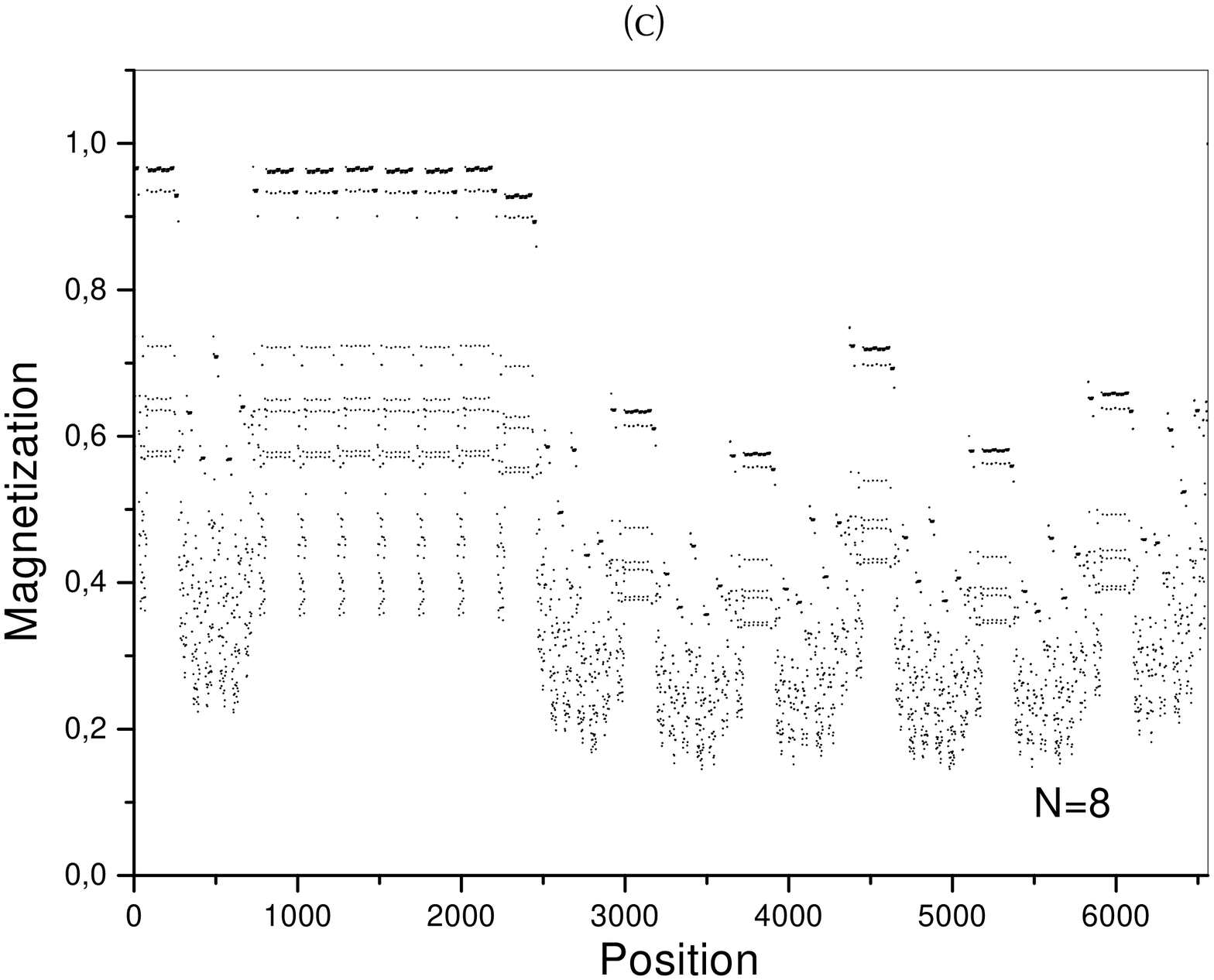,height=6cm,width=7cm}}
\caption{Magnetization profiles patterns of the model 2 for lattices 
with $N=10$ hierarchies at the criticality. (a) $R=1$ homogeneous 
system, (b) $R_{c1}=0.2706...$ and (c) $R_{c2}=6.3104...$ aperiodic systems at the fixed-points of the two-cycle attractor, respectively.}}
\end{figure}

As this alternating form of the profile pattern results from the different
properties of the two points of the cycle, it is not observed in the case of
irrelevant fluctuations. Indeed, for any value of $J_A$ and $J_B$, the
trajectory at the corresponding $T_c$ is controlled by the properties of the
critical point of the homogeneous system: the profile converges to a
definite form, with a lower symmetry than in the case of the homogeneous
system.

As previously observed for the homogeneous cases \cite{morgado90}, the high
degree of singularities suggests the multifractal analysis as a valuable
tool to investigate the structure of the magnetization profiles for the
models with $R\neq 1$. The $F(\alpha )$ functions of the measure defined by
the normalized local magnetization $\mu_i(L)=m_i(L)/\sum_jm_j(L)$ were
calculated by the method due to Chhabra and Jensen \cite{chhabra} which is
based on a one parameter family of normalized measures,

\begin{equation}
\xi _i(q,L)={\frac{\mu _i(L)^q}{\sum_j[\mu _j(L)]^q}.}  \label{eq17}
\end{equation}
In Figures 6a and 6b we show the plots for both models comparing the $%
F(\alpha )$ spectra of the aperiodic and the homogeneous system at
criticality. Both curves have the same maximum value ($F(\alpha )_{max}$=1)
since the measure has been arbitrarily assigned to the interval $[0,1]$,
that is, to the same fractal set (support) of dimension $d=1$. Therefore,
one must focus our attention to the values of the H$\ddot o$lder exponent $%
\alpha $. The minimum and the maximum values of $\alpha $ reflect,
respectively, how the measures of the most concentrated and the most
rarefied intervals scale with the box width. For the homogeneous systems the
domain of the $F(\alpha )$ spectra can be analytically calculated \cite
{coutinho92}. Moreover as was demonstrated in \cite{morgado91,coutinho92}
there is a straightforward relation between the H$\ddot o$lder exponent and
the critical exponents given by 
\begin{equation}
\alpha =d+\frac 1\nu (\beta _\alpha -\beta )  \label{eq18}
\end{equation}
where $d$ is the fractal dimension of the support and $\beta _\alpha $ is
the critical exponent associated with the average magnetization of the
subset of sites whose the measure vanishes with exponent $\alpha $, as $%
L\rightarrow \infty $. Since the most concentrated measures of both models
have finite values at the thermodynamic limit induced by the boundary
conditions we have $\beta _{\alpha _{min}}=0$ and $\alpha _{min}=d-\beta
/\nu $. Therefore $\alpha _{min}$ is solely related with the exponents
describing the critical behavior of the whole system and should follows the
rules for the universality class of the systems. 
Moreover, for the subset of sites where the measure behaves with 
the H$\ddot o$lder exponent $\overline{\alpha }=d$, the local magnetization behaves with the same critical exponent of the average magnetization of whole lattice, that is, $\beta _{\overline{\alpha }}=\beta $. This particular subset should also follow the rules of
the universality class, and therefore the point $(d,F(d))$ on the 
$F(\alpha)$ function should remains fixed when irrelevant geometric fluctuations are considered.

From Figure 6b, it becomes clear that the aperiodic systems with relevant
geometric fluctuations belong to a distinct class of universality regarding
the homogeneous system, as the $F(\alpha )$-function of the former is
shifted to higher values of $\alpha $. Moreover, we notice two distinct $%
F(\alpha )$ spectra, corresponding to distinct profiles with $R=1$ and {\bf $%
R\neq 1$.} 
\begin{figure}
\leavevmode
\vbox{%
\epsfxsize=8cm
\epsfig{file=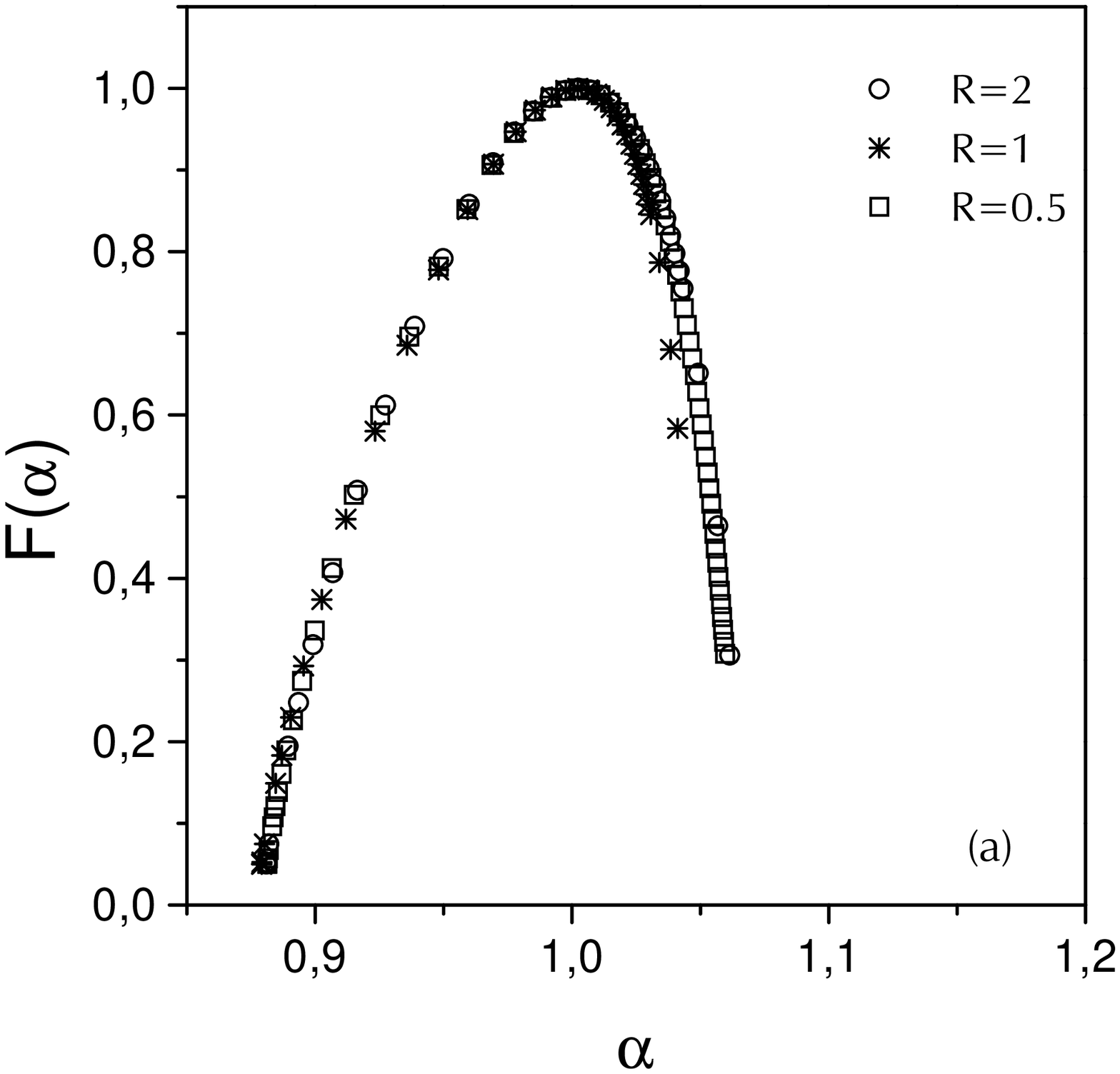,height=6cm,width=8cm}}
\leavevmode
\vbox{%
\epsfxsize=8cm
\epsfig{file=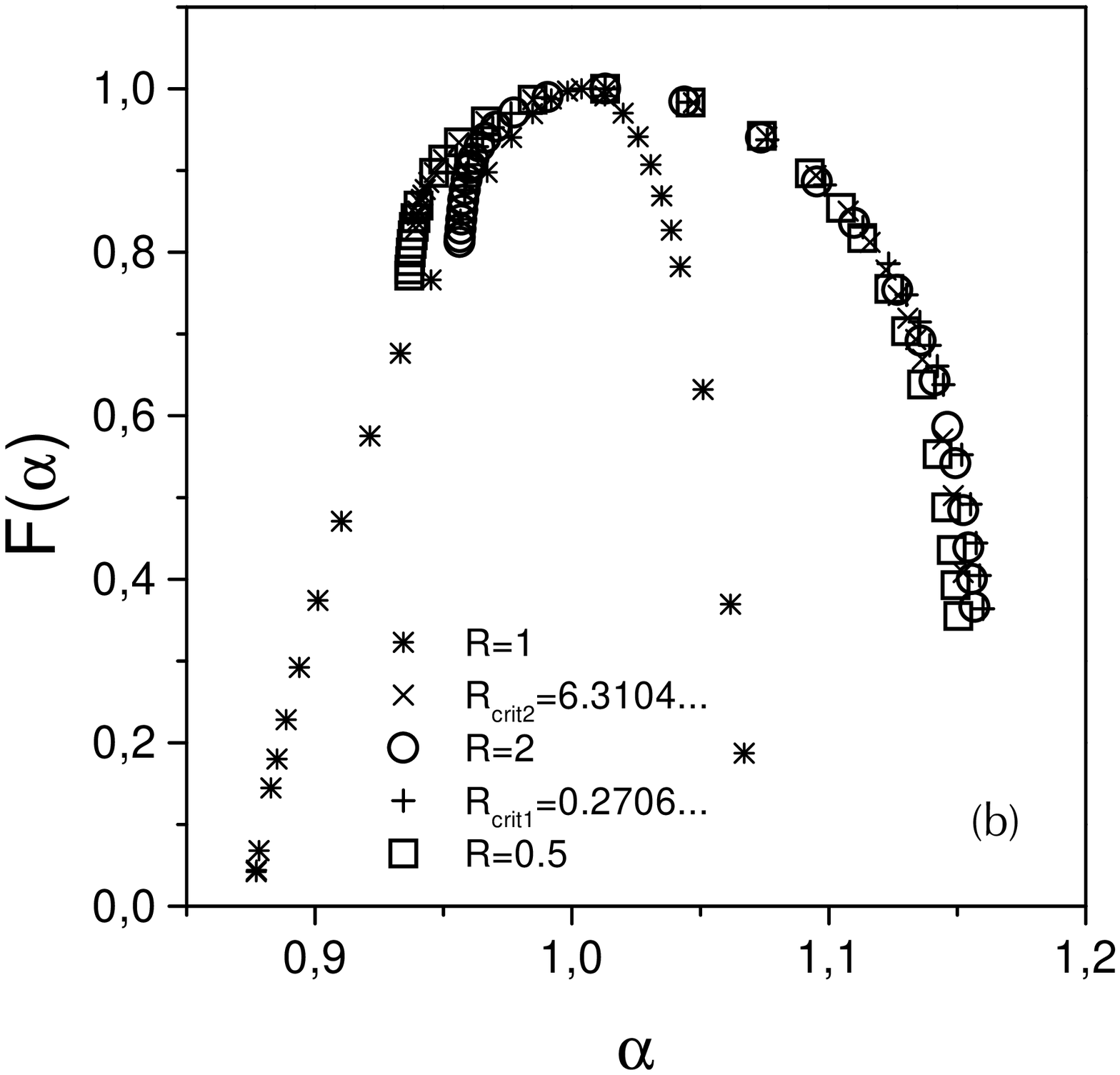,height=6cm,width=8cm}}
\caption{F($\alpha $) spectra of the normalized magnetization 
hierarchies at the criticality. (a) model 1 for lattices with 
$N=20$ and $q\in \lbrack -60,60]$. $\ast$ labels the spectrum of the corresponding homogeneous
systems, $\bigcirc$ the $R=2$ aperiodic system and $\Box$ the $R=0.5$
aperiodic system . (b) model 2 for lattices with $N=15$ and $q\in\lbrack
-40,50]$, $\ast$ labels the spectrum of the corresponding homogeneous
systems, $\bigcirc$ and $\Box$ the spectrum of the $R=2$ and $R=0.5$
aperiodic system respectively . $+$ and $\times$ label the spectrum of the
$R_{c1}=0.2706...$ and $R_{c2}=6.3104...$ fixed-points of the two-cycle
attractor, respectively.}
\label{fig6b}
\end{figure}

On the other hand, as displayed in Figure 6a the $F(\alpha )$
functions of model 1 with $R=0.5$ and $2$ suffer minor changes 
due to finite size effects when compared with the one for $R=1$. 
To handle with such differences we refine our calculations by 
doing a finite size scaling estimation of the values of $\alpha $ and $F(\alpha )$ for higher $|q|$, and
by proceeding a finite size scaling estimation of the exponent $\phi $
governing the average magnetization at the critical point $m\sim L^{-\phi }$%
,which by it turn is related with the critical exponents by $\phi =\beta
/\nu $.

In Tables \ref{tab2} and \ref{tab3}, we present these estimations for both
models respectively, comparing with exact values whenever is possible. We
notice the slight difference when the $\alpha _{min}$ of the aperiodic
system of model 1 is compared with the exact value for the homogeneous
system, indicating conservation of universality (see Table \ref{tab2}).
However, for model 2, qualitative differences are observed when the $\alpha
_{min}$ values of the aperiodic and the homogeneous models are compared. In
this latter case the values of $\alpha _{min}$ oscillates as we go from $N$
even to $N$ odd lattices. As pointed out in our former discussion, these
oscillations are a manifestation of the fact that the profile oscillates
between two different patterns dictated by the properties of each of the
points of the cycle. In Figure 7 we present the scaling for the $\alpha
_{min}$ of the F($\alpha $) function of model 2 for some values of the
fluctuation parameter $R$. For $R_{c1}$ and $R_{c2}$ these oscillations are
clearly seen as we go from $N$ even to $N$ odd lattices. For finite lattices
and finite values of $q$, the oscilations on the scaling plot introduce
significant errors bars for thevalues of the corresponding exponents.
However, as $N$ goes to infinity, we expect that $\alpha _{min}$ converges
to a unique universal limit for any value of the fluctuation parameter other
than $R=1$. 

In Table \ref{tab3}, we also compare the values of $\alpha
_{min} $ obtained within the present approach with the ones deduced from the
results for $\beta /\nu $, calculated using the transfer matrix approach
developed by one of us \cite{andrade99}. It is worth to mention that this
latter approach allows to work with higher values of $N$, ending up with
more accurate values for the exponents.
\begin{figure}
\leavevmode
\vbox{%
\epsfxsize=4cm
\epsfig{file=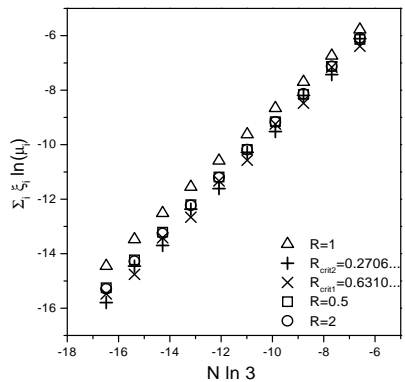,height=6cm,width=8cm}}
\caption{Scaling of  the $\alpha _{min}$ of the $F(\alpha )$-function of 
model 2 for values of $R=1$, $R=0.5$, $R=2$, $R_{c1}=0.2706...$ and 
$R_{c2}=6.3104...$. The corresponding exponents are given on table III.}
\label{fig7}
\end{figure}

\section{Summary and conclusions}

We investigate the role of the geometric fluctuations on the local
magnetization and critical properties of the ferromagnetic Ising model when
the coupling constants are defined by deterministic aperiodic sequences. Two
distinct models were considered as a matter of comparation. Model 1 (model
2) was build with irrelevant (relevant) Fibonacci like geometric
fluctuations with respect to changes in the critical properties. The models
are defined on different diamond hierarchical lattices but with then same
fractal dimension $d_F=2$. The phase diagram of both models were obtained
and carefully analyzed. The local magnetization of both models was
calculated by an exact recurrence procedure as function of the temperature
but we concentrate our analysis close to the critical point. When the
relevant (irrelevant) geometric fluctuation is concerned the pattern of the
local magnetization profile exhibit a strong (weak) change of it symmetry as
compared with the one of the corresponding homogeneous system. For the model
1, the self-similar pattern induced by the lattice topology is predominant,
as it occurs for the homogeneous system.

However, for model 2, the pattern symmetry of the homogeneous system is
strongly affected by the one dictated by aperiodic distribution of coupling
constants and the local magnetization profile resulting with a quite
different distribution of singularities. This break of symmetry is
straightforwardly related with the change in critical properties, as
revealed by the multifractal analysis of the magnetization profiles at the
criticality. The $F(\alpha )$-functions of the model 1 is quite close to the
one of the corresponding homogeneous systems for all values of domain of the
spectrum, reflecting the same critical behavior of the latter. However, the
domain of the $F(\alpha )$-functions of the model 2 is shifted to higher
values of $\alpha $, with respect to the one of the corresponding
homogeneous systems, leading to change the universality class of the model
regarding the relative values of the ferromagnetic coupling constants. The
critical exponent associated with the magnetization of model 2 were
calculated by means of the $F(\alpha )$-function as well as by scaling the
average magnetization. The results are in good agreement with the ones
obtained through the thermodynamic functions calculated using a quite
different approach previously developed by one of us. Therefore, we have
demonstrated, by means of exact numeric calculations of the local
magnetization of useful models, that as far as relevant deterministic
geometric fluctuations is concerned the resulting disorder in the
distribution of aperiodic ferromagnetic interactions leads to a strong break
of symmetry in the patterns of the local magnetization following the change
of universality class.

\acknowledgments{This work was partially supported by the Federal Brazilian
granting agencies CNPq, CAPES and FINEP (under the grant PRONEX
94.76.0004/97). One of us (S.\ C.\ ) is also grateful for the financial
support received from his local granting agency FACEPE. The authors thank S.
R. Salinas, S. T. R. Pinho and T. A. S. Haddad for helpful discussions.}
\onecolumn
\begin{table*}
\caption{Domain of the F($\alpha $) spectra for the aperiodic and
homogeneous systems for model 1. The numerical values obtained by the
scaling of $\alpha_{min}$ and $\alpha_{max}$ of the aperiodic systems were
calculated for lattices with N=10 to 22 and for $q= \pm 50$. The finite size
scaling of the average magnetization was performed for lattices with 
$L=2^N$, $N=10$ to $22$.}
\label{tab2}
\begin{tabular}{r|c|c|c|c|}
method &  & R=1 & R=2 & R=0.5 \\ \tableline 
analytic \tablenote{Ref \cite{morgado90,coutinho92}} & $\alpha_{mim}$ 
& 0.8791464205 & -- & -- \\ 
scaling \tablenote{present work} & $\alpha_{mim}$ & 0.87895(6E-5) & 0.8793(9E-4)
 & 0.876(E-3) \\ 
scaling $^{b}$ & $1-\beta/\nu$ & 0.87900(3E-5) & 0.8794(9E-4) & 0.879(E-3) \\ 
transfer matrix \tablenote{Ref. \cite{andrade99}} & $1-\beta/\nu$ & 0.87915(E-5) & -- & -- \\ 
analytic $^{a}$ & $\alpha_{max}$ & 1.04604781... & -- & -- \\ 
scaling $^{b}$ & $\alpha_{max}$ & 1.03920(7E-5) & 1.0467(6E-4) & 1.0474(5E-4)\\ 
\end{tabular}
\end{table*}
\begin{table*}
\caption{Domain of the F($\alpha $) spectra for the aperiodic and
homogeneous systems for model 2. The numerical values obtained by the
scaling of $\alpha_{min}$ and $\alpha_{max}$ of the aperiodic systems were
calculated for lattices with N=6 to 15 and for $q= \pm 40$. The finite size
scaling of the average magnetization was performed for lattices with $L=3^N$,$N=6$ to $15$.}
\label{tab3} \begin{tabular}{r|c||c|c|c|c|c|}
method &  & R=1 & R=6.3104... & R=0.2706... & R=2.0 & R=0.5 \\ \hline
analytic \tablenote{ Ref \cite{morgado90,coutinho92}} & $\alpha_{mim}$ & 0.8760357586... & -- & -- & -- & -- \\ 
scaling \tablenote{present work} & $\alpha_{mim}$ & 0.8768(5E-4) & 0.95(E-2) & 0.94(E-2) & 
0.9314(7E-4) & 0.9219(8E-4) \\ 
scaling $^{b}$ & $1-\beta/\nu$ & 0.8768(5E-4) & 0.95(E-2) & 0.94(E-2) & 
0.9301(6E-4) & 0.9178(6E-4) \\ 
transfer matrix \tablenote{Ref. \cite{andrade99}}& $1-\beta/\nu$ & 0.8760(E-4) & 0.95481(E-5) & 
0.955(E-3) & -- & -- \\ 
analytic $^{a}$ & $\alpha_{max}$ & 1.068947633.. & -- & -- & -- & -- \\ 
scaling $^{b}$ & $\alpha_{max}$ & 1.0643(6E-4) & 1.1660(5E-4) & 1.162(5E-3) & 
1.166(5E-3) & 1.160(5E-3)\\
\end{tabular}
\end{table*}
\twocolumn

\end{document}